\documentclass[aps,prb,reprint,superscriptaddress,longbibliography]{revtex4-1}

\usepackage[utf8]{inputenc}
\usepackage{graphics,color}
\usepackage{amsmath}
\usepackage{amsfonts}
\usepackage{braket}
\usepackage{natbib}
\usepackage{dsfont}
\usepackage{bm}
\setcitestyle{square}
\DeclareMathOperator{\Tr}{Tr}
\newcommand{\bs}[1]{\bm{#1}}

\begin{document}

\title{Multilevel coherences in quantum dots}

\author{Martin T. Maurer}
\affiliation{Institut f\"ur Theorie der Statistischen Physik, RWTH Aachen, 52056 Aachen,
Germany and JARA -- Fundamentals of Future Information Technology}
\author{J\"urgen K\"onig}
\affiliation{Theoretische Physik, Universit\"at Duisburg-Essen and CENIDE, 47048 Duisburg, Germany}
\author{Herbert Schoeller}
\affiliation{Institut f\"ur Theorie der Statistischen Physik, RWTH Aachen, 52056 Aachen,
Germany and JARA -- Fundamentals of Future Information Technology}

\date{\today}

\begin{abstract}
We study transport through strongly interacting quantum dots with $N$ energy levels that are weakly coupled to generic multi-channel metallic leads. In the regime of coherent sequential tunneling, where level spacing and broadening are of the same order but small compared to temperature, we present a unified, $SU(N)$-invariant form of the kinetic equation for the reduced density matrix of the dot and the tunneling current. This is achieved by introducing the concept of \textit{flavor polarization} for the dot and the reservoirs, and splitting the kinetic equation in terms of \textit{flavor accumulation}, anisotropic \textit{flavor relaxation}, as well as exchange-field- and detuning-induced \textit{flavor rotation}.
In particular, we identify the exchange field as the cause of negative differential conductance at off-resonance bias voltages appearing in generic quantum-dot models. To illustrate the notion of flavor polarization, we analyze the non-linear current through a triple quantum-dot device.
\end{abstract}

\maketitle

\section{Introduction}
The spatial confinement of electrons in quantum dots gives rise to both a charging energy and a discrete spectrum of single-particle energy levels. If two or more levels are energetically close to each other compared to their tunneling-induced broadening, coherent superpositions may form and influence the electronic transport through the quantum dots.
By coupling a spin-$\frac{1}{2}$ dot level to ferromagnetic leads (thereby forming a quantum-dot spin valve) and applying a bias voltage, the interplay of spin accumulation, relaxation, and precession gives rise to a non-equilibrium polarization of the quantum-dot spin 
\cite{QDSV_PRL,QDSV_PRB,Braig,Rudzinski,Weymann,Hornberger,Hell,Gergs,Davidovic,Hamaya,Kontos}.
Controlling transport by generating and manipulating spins is the declared goal of the field of \textit{spintronics}.

The $SU(2)$ framework for the spin degree of freedom is easily transferred to other $2$-level systems by introducing an \textit{isospin}. This includes the valley degree of freedom in the band structure of graphene and carbon nanotubes, studied in the field of \textit{valleytronics} \cite{valleytronics_1,valleytronics_2}.
Another example is given by quantum-dot Aharonov-Bohm interferometers, in which the coherent superposition of the orbital levels of two single-level quantum dots gives rise to Aharonov-Bohm oscillations of the current through the device \cite{ABexp_1,AB_2,ABexp_2}.
Furthermore, superconducting correlations in quantum dots attached to superconducting leads have been described in terms of an isospin defined by two quantum-dot states with different particle numbers \cite{super}.

In the last decades, triple quantum dots have been realized experimentally \cite{TQDexp1,TQDexp2,TQDexp3,TQDexp4}. In such structures, three instead of two states can be energetically close to each other, suggesting an $SU(3)$ framework.
Even coherences between more than three levels are realized in molecules such as benzene \cite{BenzeneHettler, BenzeneDarau}.
Common among these systems are coherence-induced transport signatures such as negative differential conductance (NDC) and complete current blockades, making them interesting for technological application in nanoelectronic devices. It is, therefore, of high interest to find a description of the complex nonequilibrium behavior of generic $N$-level dots in a unified and physically intuitive way similar to spin-valve systems. 

In this paper, we seek such a description for quantum dots with an arbitrary number $N$ of orbitals coupled to generic multi-channel metallic leads. The underlying group in this case is $SU(N)$. We will present a unified theoretical framework for the regime where the level spacing $\Delta$ and the broadening $\Gamma$ are of the same order and small compared to temperature $T$, which we refer to as the {\it coherent-sequential-tunneling regime}. It is of particular interest since it exhibits quantum coherence in weak coupling and is most easily accessible to experiments. Similar as in quantum-optics approaches \cite{Alicki}, we represent the density matrix of the dot by a real vector, which we refer to as the \textit{flavor polarization} of the dot. In addition, we define also a set of flavor polarizations for the reservoirs, which is crucial to understand the NDC physics induced by quantum coherence. We show that the kinetic equations governing the dot dynamics can be cast in a universal, $SU(N)$-invariant form containing terms that describe dot-flavor accumulation, relaxation and rotation, suggesting the term {\it flavortronics} to describe transport through $N$-level quantum dots. A central result of our work is the identification of flavor rotations as the generic cause of NDC at off-resonance bias voltages. We illustrate this and the general usefulness of the flavor-polarization formalism by analyzing the I-V-characteristic of a triple-dot setup.

\section{Model}
We consider $N$ spinless quantum-dot orbitals with strong Coulomb interaction that are weakly coupled to multi-channel metallic leads. 
The total Hamiltonian is given by $H  = H_{\text{D}}+ H_{\text{res}} + H_{\text{T}}$. 
For convenience, we work in a basis where the single-particle part of the dot Hamiltonian is already diagonalized. Including the interaction, the dot is described by $H_{\text{D}} = H_{\text{D}}^0 + H_{\text{D}}^{\text{int}} = \sum_{i=1}^N \epsilon_i c_i^{\dagger} c_i + U \sum_{i<i^{\prime}}c^{\dagger}_i c_i c^{\dagger}_{i^{\prime}}c_{i^{\prime}}$. 
The average level position is defined by $\epsilon=\sum_{i=1}^N \epsilon_i/N$, the detunings by $\Delta_{ij} = \epsilon_i - \epsilon_j$.
For large Coulomb interaction, $U\rightarrow \infty$, only the empty and the singly-occupied dot configurations are allowed.
The leads $H_{\text{res}}= \sum_{r=1}^{N_\text{res}} H_r$ with $H_r  = \sum_{k\nu} \epsilon_{rk\nu} a^{\dagger}_{rk\nu} a_{rk\nu}$ are modeled as reservoirs of noninteracting electrons with temperature $T$ and chemical potential $\mu_r$.
The channel index $\nu=1,\ldots,N_{\text{ch}}$ accounts for different bands, and the quantum number $k$ labels the energy eigenstates in each band. 
The reservoir density of states $\rho_r(\omega) = \rho_0 \Lambda^2/\left[(\omega-\mu_r)^2 + \Lambda^2 \right]$ contains a high-energy cutoff $\Lambda$ ensuring convergence of appearing integrals.
Tunneling between dot and leads is described by $H_{\text{T}} = \sum_{r\nu i} t^r_{\nu i} a^{\dagger}_{rk\nu} c_i + \text{h.c.}$, with energy-independent tunneling amplitudes $t^r_{\nu i}$. 
The latter enter the $N\times N$ hermitian, positive semidefinite hybridization matrices $\Upsilon^r$ with matrix elements $\Upsilon^r_{ij} = 2\pi\rho_0\sum_{\nu} (t^r_{\nu i})^* t^r_{\nu j}$.
The tunnel-coupling strength to reservoir $r$ is characterized by $\Gamma^r = \Tr \Upsilon^r/N$, and the total tunneling strength by $\Gamma = \sum_r \Gamma^r$.
We set $e=\hbar = k_B = 1$ throughout this paper.

\section{Flavor representation of the quantum-dot state}
Since the infinite charging energy limits the number $N_e$ of electrons in the quantum dot to $0$ and $1$, the Hilbert space of the quantum-dot states is $N+1$-dimensional with basis states $\ket{0}$ for an empty quantum dot and $\ket{i}$ for an electron occupying level $i = 1,2,\ldots,N$.
As a result, the reduced density matrix $\rho = \rho_{N_e=0} + \rho_{N_e=1}$ of the quantum dot can be decomposed into a part $\rho_{N_e=0} = P_0\ket{0}\bra{0}$ describing the empty quantum dot (with probability $P_0$) and a part $\rho_{N_e=1}$ for single occupation (with probability $P_1 = \Tr  \rho_{N_e=1} = 1 - P_0$).
The latter is a $N\times N$ hermitian, positive semidefinite matrix that can be decomposed into the identity matrix $\mathds{1}_N$ and a set of $s_N=N^2-1$ traceless generators $\lbrace{\lambda_a\rbrace}$ of $SU(N)$, which are normalized such that
$\Tr(\lambda_a \lambda_b) = 2\delta_{ab}$, $[\lambda_a, \lambda_b]_- = 2i\sum_c f_{abc}\lambda_c$, and 
$[\lambda_a, \lambda_b]_+ = \frac{4}{N}\delta_{ab} + 2 \sum_c d_{abc} \lambda_c$,
with real constants $f_{abc}$ and $d_{abc}$ forming a totally antisymmetric and a symmetric tensor, respectively
\footnote{The numerical values of these constants depends on the chosen set of generators. A straightforward choice are the generalized Gell-Mann matrices \cite{Hioe}.}. As a result \cite{Byrd,Kimura}, the density matrix for single occupation,
\begin{align}
	\rho_{N_e=1} = \frac{1}{N}\left(P_1 \mathds{1}_N + c_N\,\bs{g}\cdot \bs{\lambda}\right) \:, \label{EQ:rho_N}
\end{align}
with $\bs{g}\cdot \bs{\lambda} = \sum_a g_a \lambda_a$ and $c_N = \sqrt{N(N-1)/2}$, is parametrized by the probability $P_1$ of single occupation and the components $g_a$ of an $s_N$-dimensional real vector $\bs{g}$, referred to as \textit{flavor polarization} of the dot.
Semi-positivity of $\rho_{N_e=1}$ implies $\Tr \rho_{N_e=1}^2 \leq P_1^2$, which yields $|\bs{g}|\leq P_1$, i.e., the normalization is chosen such that $|\bs{g}|=1$ describes maximal flavor polarization. The $s_N$-dimensional flavor-polarization vector $\bs{g}$ generalizes the three-dimensional spin-polarization vector in the case of a spinful quantum-dot level for $N=2$ to any number $N$ of quantum dot levels. We note that for $N>2$, flavor polarization is fundamentally different from angular momentum $(N-1)/2$, as the latter is described in terms of the $N$-dimensional representation of the three generators of $SU(2)$, and not of the $s_N$ generators of $SU(N)$.

The dot flavor polarization carries the information about the mixture and superpositions of dot states contained in the density matrix. The modulus $|\bs{g}|$ is a measure for the purity in the one-particle sector, defined as $\gamma = \Tr[(\rho_{N_e=1}/P_1)^2] = [1+(N-1)(|\bs{g}|/P_1)^2]/N$ \footnote{The scaling by $P_1$ in this definition ensures that the purity takes the usual values $\gamma \in [1/N, 1]$.}. 

Thus, maximal flavor polarization $|\bs{g}_1|=1$ corresponds to a pure state in which, in a properly chosen basis, one of the $N$ dot levels is occupied with probability $1$.
All mixed or pure states with this specific dot level being empty are described by flavor-polarization vectors $\bs{g}_2$ that satisfy the condition $\bs{g}_1 \cdot \bs{g}_2=-1/(N-1)$. In contrast, vanishing flavor polarization corresponds to the maximally mixed state.

The notion of an $s_N$-dimensional flavor polarization vector is not only needed for the dot but also for each reservoir. The reservoir flavor polarization $\bs{n}^r$ (with $|\bs{n}^r| \le 1$) is defined by the decomposition 
\begin{align}
	\Upsilon^r = \Gamma^r(\mathds{1}_N+c_N\,\bs{n}^r \cdot \bs{\lambda}) \label{EQ:hyb_mat}
\end{align}
of the hybridzation matrix, i.e., $\Gamma^r$ and $\bs{n}^r$ contain all microscopic details of the tunnel coupling. Full polarization, $|\bs{n}^r| = 1$, occurs when all channels couple to the same dot state, while vanishing polarization, $\bs{n}^r = 0$, corresponds to $N$ channels that are coupled with equal strength to a different one of the $N$ dot levels each.

To determine the components of $\bs{g}$ and $\bs{n}^r$ for given density and hybridization matrices, we make use of the orthogonality of the generators to arrive at $g_a = N\Tr(\rho_{N_e=1}\lambda_a)/(2c_N)$ and $\Gamma^r n^r_a = \Tr(\Upsilon^r \lambda_a)/(2 c_N)$. 
Finally, we remark that only a subset of the vectors $\bs{g}$ or $\bs{n}^r$ in the $s_N$-dimensional unit sphere describe flavor polarization, i.e., correspond to a (positive semidefinite) density or hybridization matrix \cite{Byrd,Kimura,Jakobczyk}.

\section{Kinetic Equation}
The quantum-dot state, including its flavor polarization, is described by the reduced density matrix $\rho$ with matrix elements $\rho_{\chi \chi'} = \big \langle \ket{\chi'}\bra{\chi} \big\rangle$.
The natural basis states $\ket{\chi}$ are the empty dot $\ket{0}$ and single occupation $\ket{i}$ of level $i=1,2,\ldots,N$.
The diagonal entries $\rho_{\chi\chi}$ are the probabilities to find the dot in state $\ket{\chi}$, while the off-diagonals $\rho_{ij}$ describe coherences between level $i$ and $j$. In the weak-coupling and Markov regime, $t^{-1},\Gamma \ll T$, the kinetic equations of $\rho_{\chi\chi'}$ read
\begin{align}
	\frac{d}{dt}\rho_{\chi\chi'} = -i(\epsilon_\chi-\epsilon_{\chi'})\rho_{\chi\chi'} +\sum_{\eta\eta^{\prime}}W_{\chi\chi',\eta\eta'} \rho_{\eta\eta'} \:. \label{EQ:kin_eq}
\end{align}
The generalized transition matrix elements $W_{\chi\chi',\eta\eta'}$ in Liouville space, represented as irreducible diagrams on the Keldysh contour, are calculated up to first order in $\Gamma$ employing a real-time diagrammatic technique presented in \cite{ResTun,BosTun}, see App.\ \ref{AppA} for details. The current $I_r$ from the dot into reservoir $r$ can then be calculated from $\rho$ and a partial selection of diagrams. 

In the {\it coherent-sequential-tunneling regime}, $|\Delta_{ij}| \lesssim \Gamma$, we express the kinetic equations in terms of the flavor polarization in a coordinate-free form that makes the $SU(N)$ invariance explicit, see App.\ \ref{AppD}. This is done by reading (\ref{EQ:kin_eq}) as a matrix equation, inserting the flavor decompositions (\ref{EQ:rho_N}) and (\ref{EQ:hyb_mat}) for each appearing density and hybridization matrix, and using the relations $P_1=\sum_{i=1}^N \rho_{ii}$ and $g_a = N\Tr(\rho_{N_e=1} \lambda_a)/(2c_N)$.
We find
\begin{align}
	\frac{dP_1}{dt} \!&= \sum_r \Gamma^r\left[ N f_r^+(\epsilon) P_{0} -  f_r^-(\epsilon)\left(P_1 + (N-1) \boldsymbol{n}^r\cdot \boldsymbol{g}\right)\right] 						\label{EQ:dP_00/dt}\:, 
\end{align}
for the total-occupation number and
\begin{align}
	\frac{d\bs{g}}{dt} &= \left(\frac{d\bs{g}}{dt}\right)_{\text{acc}} + \left(\frac{d\bs{g}}{dt}\right)_{\text{rel}} + \left(\frac{d\bs{g}}{dt}\right)_{\text{rot}}\:,\label{EQ:kineqsplit}\\
	\left(\frac{d\bs{g}}{dt}\right)_{\text{acc}} &= \sum_r \Gamma^r \left[ N f_r^+(\epsilon)P_0 - f_r^-(\epsilon) P_1 \right] \bs{n}^r\: ,\label{EQ:kineqsplit_acc}\\
	\left(\frac{d\bs{g}}{dt}\right)_{\text{rel}} &= - \sum_r \Gamma^r f_r^-(\epsilon) (\bs{g} + \bs{n}^r * \bs{g}) \:,\label{EQ:kineqsplit_rel} \\
	\left(\frac{d\bs{g}}{dt}\right)_{\text{rot}} &=  \bs{B}_{\text{tot}}\wedge \bs{g} \label{EQ:kineqsplit_rot}\:,
\end{align}
for the flavor polarization.
Here, $f_r^+(\epsilon) = 1/[\exp(\beta(\epsilon-\mu_r))+1]$ is the Fermi function with $\beta = 1/T$, $f_r^-(\epsilon) = 1-f_r^+(\epsilon)$, $\bs{B}_{\text{tot}} = \bs{B} + \bs{B}_{\text{ex}}$, $\bs{B} = \Tr(H_{\text{D}}^0 \bs{\lambda})/c_N$, $\bs{B}_{\text{ex}}=\sum_r \bs{B}_{\text{ex}}^r$, and
\begin{align}
    \bs{B}^r_{\text{ex}} = \frac{\Gamma^r}{\pi}\left[\Re \: \psi\left(\frac{\pi+i\beta(\mu_r-\epsilon)}{2\pi}\right) - \psi \left(\frac{\pi+\beta \Lambda}{2\pi}\right)\right] \bs{n}^r \:, \label{EQ:Bex}
\end{align}
with the digamma function $\psi$. The star/wedge products $(\bs{x} * \bs{y})_a = c_N\sum_{bc} d_{abc} x_b y_c$ and $(\bs{x} \wedge \bs{y})_a = c_N\sum_{bc} f_{abc} x_b y_c$ are straightforward generalizations of those defined for the $SU(3)$ case in \cite{SU3Pramana} and respect the $SU(N)$ invariance. The equation for $P_0$ follows simply from $dP_0/{dt} = -dP_1/{dt}$.

The kinetic equations essentially generalize those for the spin in a quantum-dot spin valve \cite{QDSV_PRB} to arbitrary flavor number $N$.
The equations show that dot occupation $P_1$ and flavor polarization $\bs{g}$ are coupled. 
The scalar product $\bs{n}^r \cdot \bs{g}$ reflects how strongly the dot electron couples to reservoir $r$. 
This affects the rate of tunneling processes from the dot into $r$, see Eq.~(\ref{EQ:dP_00/dt}). 

We have split the equation for $d\bs{g}/dt$ into three parts. 
The first part, (\ref{EQ:kineqsplit_acc}), describes \textit{flavor accumulation} due to tunneling between dot and flavor-polarized reservoirs.
For each reservoir $r$, the contribution to flavor accumulation is proportional to $\bs{n}^r$. 

The second term, (\ref{EQ:kineqsplit_rel}), describes \textit{flavor relaxation}. 
It can be written as $\left(d\bs{g}/dt\right)_{\text{rel}} = - \sum_r \Gamma^r f_r^-(\epsilon) D^r \bs{g}$ by introducing the matrix $D^r$ with matrix elements $D^r_{ac} = \delta_{ac} + c_N \sum_{b} d_{abc} n^r_b$.
Because $D^r$ is positive semidefinite (see App.\ \ref{AppC}),
the relaxation term always reduces the modulus of the flavor polarization, $\left(d|\bs{g}|/dt\right)_{\text{rel}} \le 0$.
The matrix $D^r$ differs from the identity matrix, which makes flavor relaxation \textit{anisotropic} \footnote{An exception is the $SU(2)$ case, where the $d$'s are vanishing, which makes spin relaxation in quantum-dot spin valves \textit{isotropic} \cite{QDSV_PRB}.}.

The last term, (\ref{EQ:kineqsplit_rot}), describes \textit{flavor rotation}. 
It can be rewritten as $(d\bs{g}/dt)_{\text{rot}} = F \bs{g}$ by introducing the matrix $F$ with matrix elements 
$F_{ac} = c_N\sum_{b} f_{abc} B_{\text{tot},b}$.
Due to $f_{abc}=-f_{cba}$, $F$ is skew symmetric and, therefore, generates an $s_N$-dimensional rotation \footnote{Similar rotations of coherence vectors have been discussed in the context of quantum optics \cite{Hioe} and open quantum systems in general \cite{Alicki}}.
Two mechanisms lead to flavor rotation. The detuning-induced part $\bs{B}$ generalizes the Zeeman-field induced spin rotation in the $SU(2)$ case.
The contribution $\bs{B}_{\text{ex}}$ is induced by virtual tunneling of quantum-dot electrons into the flavor-polarized reservoirs and back.
We call $\bs{B}_{\text{ex}}$ an \textit{exchange field}, in analogy to the one leading to Larmor precession of the spin in quantum-dot spin valves \cite{QDSV_PRL,QDSV_PRB}. Besides its dependence on the reservoir flavor polarizations, its magnitude can be controlled via bias voltage, level positions, and coupling strengths, see Eq.\ (\ref{EQ:Bex}). As the term in brackets in Eq.\ (\ref{EQ:Bex}) is nonzero in the wide-band limit of large $\Lambda$, the individual reservoir exchange fields are nonzero for polarized reservoirs, and the total exchange field $\bs{B}_{\text{ex}}$ can only vanish in highly symmetric setups where different $\bs{B}^r_{\text{ex}}$ cancel.

The flavor polarization affects transport through the quantum dot.
In the coherent-sequential-tunneling regime, $|\Delta_{ij}| \lesssim \Gamma$, the current into reservoir $r$ is
\begin{align}
	I_r = \Gamma^r \left[ -N f_r^+(\epsilon)P_0 + f_r^-(\epsilon)\left(P_1 + (N-1)\bs{n}^r \cdot \bs{g}\right)\right] \:.  \label{EQ:SUN_current}
\end{align}
In the special case of a singly-occupied dot, $P_1=1$, and a flavor polarization $\bs{g}$ satisfying $\bs{n}^r \cdot \bs{g}=-1/(N-1)$, no current flows into the reservoir.
This \textit{flavor blockade} appears since the states corresponding to $\bs{g}$ decouple from the reservoir.

The kinetic equations (\ref{EQ:dP_00/dt})--(\ref{EQ:kineqsplit_rot}) and the current formula (\ref{EQ:SUN_current}) are the main results of our paper. They provide an intuitive picture of the dot dynamics and the electronic transport in terms of the flavor polarization. We emphasize the special role of the exchange field (\ref{EQ:Bex}). 
Its dependence on the chemical potentials is responsible for the NDC at off-resonance bias voltages, where all Fermi functions are constant.
The precise mechanism is discussed
below for the simple example of a triple quantum dot, but the same reasoning applies to any setup with $N$ levels in the coherent-sequential-tunneling regime.
While current blockades due to coherence effects and resulting NDC have been widely studied \cite{NDC1,DS1,DS2,DS5,DS6,DS8,DS3,DS7,NDC2,DS9,DS4,Niklas}, this intuitive explanation of off-resonance NDC for generic $N$-level setups closes a gap in the literature.

In the opposite {\it incoherent-sequential-tunneling regime of large detunings, $|\Delta_{ij}|\sim T\gg\Gamma$, }the coherences can be neglected, and both contributions to the rotation term drop out. In that case, the kinetic equations simplify to the standard Fermi's golden rule rate equations, $d\rho_{ii}/dt = \sum_r \left[ \Upsilon^r_{ii} f_r^+(\epsilon_i) \rho_{00} - \Upsilon^r_{ii} f_r^-(\epsilon_i) \rho_{ii} \right]$ and $d\rho_{00}/dt = - \sum_{i=1}^N d\rho_{ii}/dt$, as well as $I_r = \sum_{i} \left[ -\Upsilon^r_{ii} f_r^+(\epsilon_i) \rho_{00} + \Upsilon^r_{ii} f_r^-(\epsilon_i) \rho_{ii} \right]$.

Let us briefly consider the general case, where the levels are arranged in multiple groups of close-lying energies. This case can be treated straightforwardly by the formalism. All isolated levels enter the master equations via the Fermi's golden rule rate equations. Regarding groups of at least two close-lying levels, flavor equations must be set up for each group, defining adequate flavor polarizations from the projections of the density and hybridization matrices onto the subspace of states included in the group.

Finally, we remark that an additional spin degeneracy of the quantum-dot levels can be easily taken into account without doubling $N$. 
All presented formulas remain valid once $P_0=\rho_{00}$ appearing on the r.h.s.\ is multiplied with a factor of $2$, while $\rho_{ij}$ is understood as $\sum_\sigma \rho_{i\sigma, j\sigma}$, i.e., spin affects the results only quantitatively.
In the following example, we assume spin-less electrons.

\section{Example}
\begin{figure}[!t]
\includegraphics{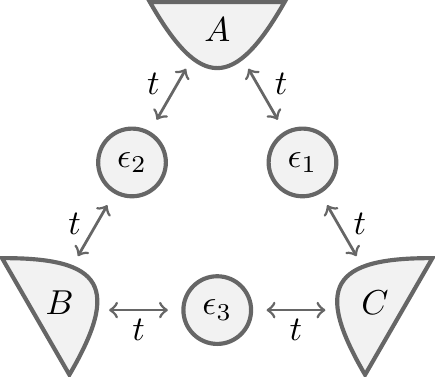}
	\caption{Three reservoirs are coupled to two levels each. Nonzero tunnel couplings are chosen real and equal. A bias voltage $V$ is applied such that $\mu_{B} = \mu_{C} = -\mu_A = V/2$. \label{Fig:1}}
\end{figure}

We illustrate the usefulness of the concept of flavor polarization by analyzing the current through the triple-dot setup shown in Fig.~\ref{Fig:1}.
Each of the three reservoirs $r=A,B,C$ couples symmetrically to two dot levels, such that $\Gamma^r = \Gamma/3$, and accommodates one channel only, which implies maximal flavor polarization ($|\bs{n}^r|=1$).
We choose the standard Gell-Mann matrices \cite{GellMann} (see App.\ \ref{AppE} for a list) as the generators of $SU(3)$.
Then, the explicit flavor-polarization vectors are given by $\bs{n}^A = (\sqrt{3}/2,0,0,0,0,0,0,1/2)$, $\bs{n}^B = (0,0,-\sqrt{3}/4,0,0,\sqrt{3}/2,0,- 1/4)$, and $\bs{n}^C = (0,0,\sqrt{3}/4,\sqrt{3}/2,0,0,0,- 1/4)$.
The chemical potentials are set to $\mu_{B} = \mu_{C} = -\mu_A = V/2$, i.e., leads $B$ and $C$ can be combined into a single lead $BC$ with flavor polarization $\bs{n}^{BC} = (\bs{n}^B + \bs{n}^C)/2$ and coupling strength $\Gamma^{BC}=2\Gamma/3$.
Using the flavor framework, we will be able to explain NDC and current blockades due to coherence effects (similar as reported in Refs.~\cite{NDC1,DS1,DS2,DS5,DS6,DS8,DS3,DS7,NDC2,DS9,DS4,Niklas}) in terms of flavor blockade and its lifting by flavor rotation.

\begin{figure}[!t]
\includegraphics{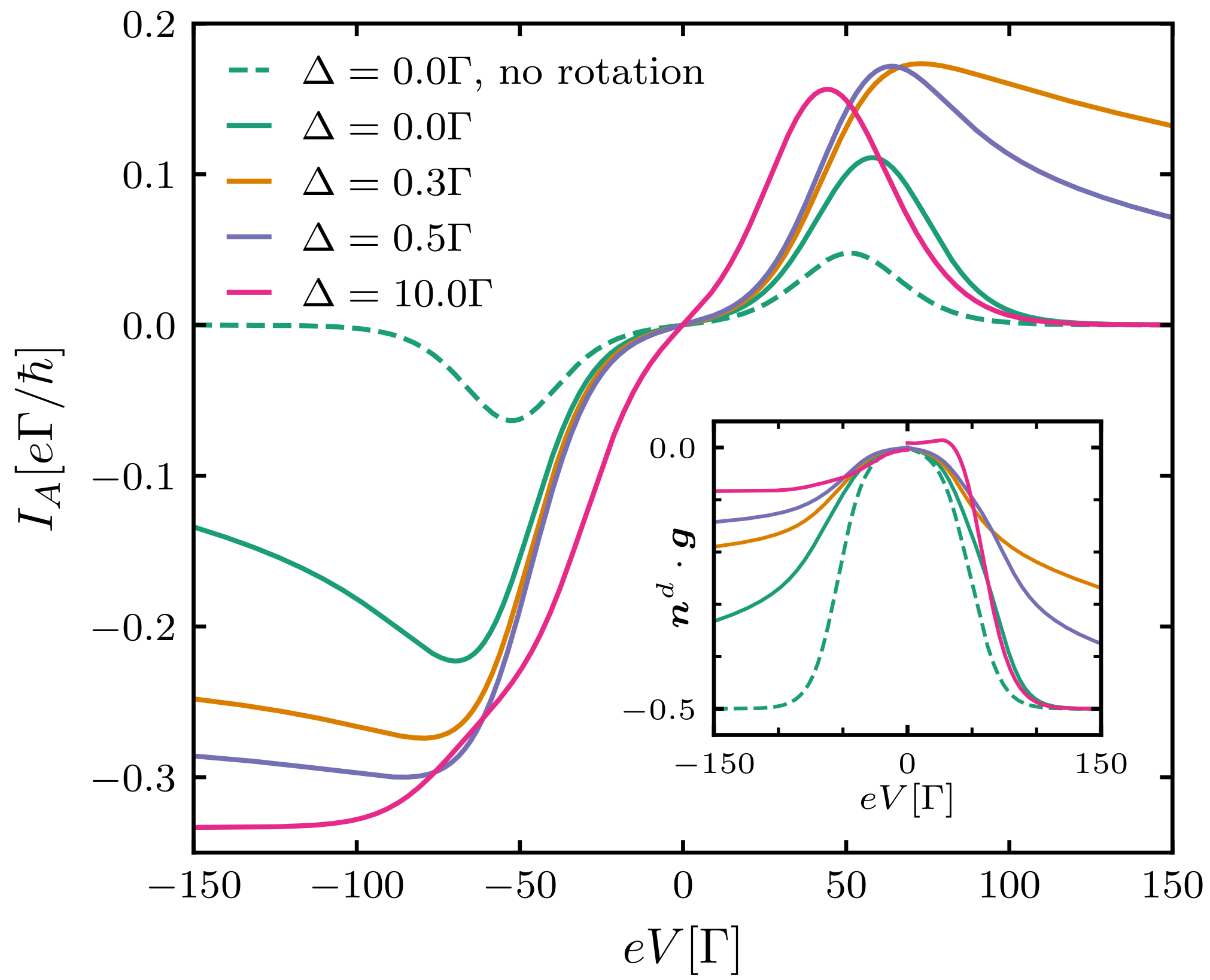}
	\caption{Current-voltage characteristics for $\epsilon= 25\Gamma$ and detunings $\Delta_{13}=-\Delta_{23}=\Delta$ calculated with the flavor equations (\ref{EQ:dP_00/dt}-\ref{EQ:SUN_current}) (for $\Delta=0, 0.3\Gamma, 0.5 \Gamma$) and with Fermi's golden rule (for $\Delta=10\Gamma$). For the dashed curve, the rotation term (\ref{EQ:kineqsplit_rot}) has been omitted by hand. 
	The inset shows the scalar product of drain and dot flavor polarization. Further parameters are $T=5 \Gamma$, $\Lambda=1000 \Gamma$. \label{Fig:2}}
\end{figure}

In Fig.~\ref{Fig:2}, we show the current into reservoir $A$ for an average dot-level energy of $\epsilon = \epsilon_3=25\Gamma$ and symmetric detunings $\epsilon_{1/2}=\epsilon\pm\Delta$ as a function of bias voltage $V$. 
We find the expected increase in current as the chemical potentials approach the dot level energies. At larger voltages,
the current exhibits signatures of quantum coherence for detunings of the order of $\Gamma$.
For $V<0$, lead $BC$ is the drain electrode, $\bs{n}^d=\bs{n}^{BC}$. At large voltages and zero detuning, a full suppression of the current is obtained when omitting the rotation term (\ref{EQ:kineqsplit_rot}) by hand (dashed line). In this case, the steady-state flavor polarization becomes $\bs{g} = (1,0,0,-1,0,-1,0,0)/\sqrt{3}$, which corresponds to the occupation of the dark state $\ket{\psi_{123}}=(\ket{1}+\ket{2}-\ket{3})/\sqrt{3}$ that decouples from the drain, i.e., the flavor-blockade conditions $P_1=1$ and $\bs{n}^{d} \cdot \bs{g} =-1/(N-1) = -1/2$ are satisfied. The blockade is partially lifted when the exchange-field- and detuning-induced flavor rotation is taken into account (see solid lines and inset), as they rotate the flavor polarization away from the blocking orientation. The magnitude of the exchange field falls off like $|\ln(|V|/2\Lambda)|$ at large voltages, which explains the observed NDC. Since away from resonance, $|\mu_r-\epsilon|\gg T$, all Fermi functions are either 0 or 1, the voltage dependence of $\bs{B}_{\text{ex}}$ is the \textit{sole} cause of the NDC appearing here. 
While the perfect blockade in the absence of flavor rotation is not a generic feature, this reasoning actually applies to NDC in any multilevel-dot model: The exchange field rotates the flavor polarization into an orientation that increases $\bs{n}^d \cdot \bs{g}$, i.e., couples more strongly to the drain, and an NDC appears because $|\bs{B}_{\text{ex}}|$ decays with increasing voltage. 

Returning to the model at hand, for large detuning (pink line), coherences are absent. This implies that flavor rotations vanish, but as the dark state $\ket{\psi_{123}}$ is a coherent superposition, it is not occupied to begin with, and the current is not suppressed.

For $V>0$, lead $A$ becomes the drain electrode, $\bs{n}^d=\bs{n}^{A}$. Our maximally symmetric model shows (nongeneric) striking current signatures here, which can easily be explained in the flavor framework. 
At zero detuning (green line) the flavor polarization is $\bs{g}= (-\sqrt{3}/2,0,0,0,0,0,0,1/2)$, which corresponds to the occupation of the dark state $\ket{\psi_{12}}= (\ket{1}-\ket{2})/\sqrt{2}$ and satisfies the flavor-blockade conditions $P_1=1$ and $\bs{n}^{d} \cdot \bs{g} = -1/2$. 
In contrast to $V<0$, flavor rotations do not restore the current since they cannot affect the dark state, as $\bs{n}^A \wedge \bs{g} =0$ and $\bs{n}^{BC} \wedge \bs{g} = 0$. This changes with small $|\Delta|$, where the flavor is rotated by the detuning-induced field $\bs{B}$. The resulting flavor is then affected by exchange-field-induced rotations, and similar as for $V<0$, off-resonance NDC appears because of the $V$-dependence of the exchange field. For large detuning, current is suppressed again since once an electron enters level $3$, it cannot leave anymore. However, compared to zero detuning, the physics involved is fundamentally different since the blockade can be understood in a simple Fermi's golden rule approach.

\section{Conclusion}
We have introduced the concept of flavor polarization for the dynamics of $N$ quantum-dot levels in the coherent-sequential-tunneling regime. 
The significance of the kinetic equations presented in this paper is threefold: Firstly, they constitute a unifying description of multilevel quantum dots. Secondly, they allow for an intuitive interpretation of the dynamics in these systems in terms of accumulation, relaxation, and rotation of a flavor-polarization vector. Thirdly, they isolate the entire bias-voltage dependence beyond the Fermi functions in a single term---the exchange field---which reveals flavor rotations as the origin of negative differential conductances in off-resonance regimes.

Our framework can straightforwardly be generalized to arbitrary occupations by introducing several dot flavor polarizations \cite{in_prep}. Furthermore, it will be also very useful for strong dot-lead coupling by taking higher-order tunneling processes into account using, e.g., real-time renormalization group methods \cite{RTRG}, where broadening and renormalization effects influence the resonance lineshapes \cite{lindner_etal_prb_19}, and the Kondo effect occurs in the cotunneling regime \cite{Goet,SU3_Kondo_Lopez,SU3_Kondo_Lindner,Arnold,Paaske}.

\begin{acknowledgments}
We thank R. Harlander, C. Lindner, and S. Siccha for fruitful discussions. This work was supported by the Deutsche Forschungsgemeinschaft via RTG 1995 and CRC 1242 (Project No. 278162697). Simulations were performed with computing resources granted by RWTH Aachen University under project thes0595. \end{acknowledgments}

\appendix
\section{Diagrams \label{AppA}}
The generalized transition matrix elements $W_{\chi \chi^{\prime},\eta \eta^{\prime}}$ are represented as irreducible diagrams on the Keldysh contour. The physical time axis runs from left to right, while the Keldysh contour runs from left to right and then back again. 
The rules for the evaluation of a diagram $W_{\chi \chi^{\prime},\eta \eta^{\prime}}$ to first order in the tunneling strength $\Gamma$ are:
\begin{enumerate}
\item Draw all topologically different diagrams with states $\eta$, $\eta^{\prime}$ to the left and $\chi$, $\chi^{\prime}$ to the right. Assign dot states and their energies to all Keldysh contour elements between vertices representing the tunneling Hamiltonian. Vertices are connected in pairs by directed tunneling lines that carry a reservoir index $r$ and tunneling energy $\omega$. A first-order diagram contains one tunneling line connecting two vertices on the far left and far right of the diagram.
\item Each segment between vertices gives a factor 
$1/(E+i0^+)$, with $E$ being the difference of all energies going to the left minus all energies going to the right, including the tunneling line energy.
\item A tunneling line with index $r$ going from a vertex where a dot state $i$ is annihilated to a vertex where a dot state $j$ is created implies a factor $\Upsilon^r_{ji} \bar{\rho}_r(\omega) f_r^{\pm}(\omega)/(2\pi)$, where $\bar{\rho}_r(\omega) = \rho_r(\omega)/\rho_0$, and $f_r^+(\omega)$ is to be taken if the line goes backward w.r.t.\ the Keldysh contour and $f_r^-(\omega)$ if it goes forward.
\item Assign a total prefactor $(-i)$ and for each vertex on the lower contour a prefactor $-1$.
\item Sum over internal indices and integrate over the tunneling energy $\omega$.
\end{enumerate}
\begin{figure}[b]
\centering
\includegraphics{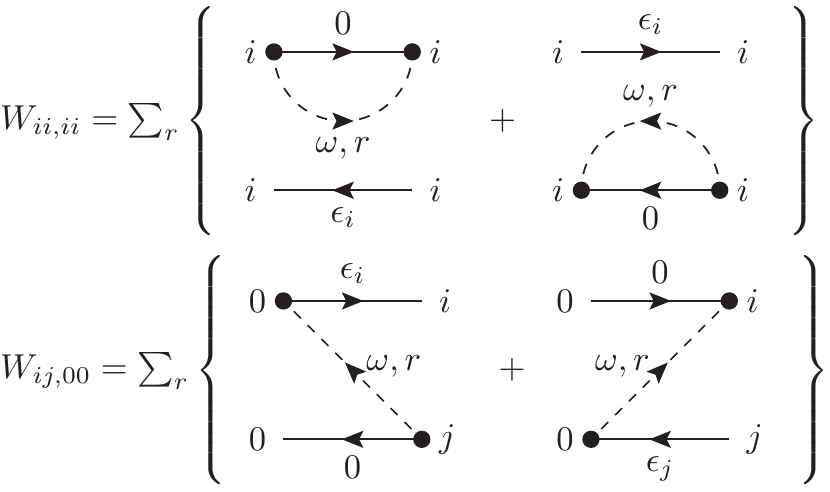}
\caption{Diagrams for two generalized transition matrix elements. \label{Fig:examples}}
\end{figure}

As an example, Fig. \ref{Fig:examples} shows the diagrams for two generalized transition matrix elements, with $i,j=1,\ldots,N$ labeling a dot level. According to the above rules, their values are in the limit of large $\Lambda$:
\begin{widetext}
\begin{align}
W_{ii,ii} &= -i\sum_r \frac{\Upsilon^r_{ii}}{2\pi} \int d\omega \: \bar{\rho}_r(\omega) f_r^-(\omega)\left(\frac{1}{\epsilon_i - \omega +i0^+} + \frac{1}{\omega - \epsilon_i + i0^+}\right)\nonumber \\
&= - \sum_{r} \Upsilon^{r}_{ii}f_{r_i}^-(\epsilon_i) \: ,\\
W_{ij,00} &=i\sum_r \frac{\Upsilon^r_{ij}}{2\pi} \int d\omega \: \bar{\rho}_r(\omega) f_r^+(\omega)\left(\frac{1}{\omega - \epsilon_i + i0^+} + \frac{1}{\epsilon_j - \omega + i0^+}\right) \nonumber\\
&=\sum_r\frac{\Upsilon^r_{ij}}{2}\Big[\left(f_r^+(\epsilon_i) + f_r^+(\epsilon_j)\right) +i \big(\Omega_r(\epsilon_i) - \Omega_r(\epsilon_j)\big)\Big] \: ,
\end{align}
\end{widetext}
with $\Omega_r(\epsilon_i)= \frac{1}{\pi}\left[\Re \: \psi\left(\frac{1}{2}+i\frac{\beta}{2\pi}(\mu_r-\epsilon_i)\right) - \psi \left(\frac{1}{2}+\frac{\beta \Lambda}{2\pi}\right)\right]$, where $\psi$ is the digamma function.

The current into reservoir $r$ 
reads to first order:
\begin{align}
I_r = \sum_{\chi \eta \eta^{\prime}} \sum_m mW_{\chi \chi,\eta \eta^{\prime}}^{rm} \rho_{\eta \eta^{\prime}} \: .
\end{align}
Here, $W_{\chi \chi,\eta \eta^{\prime}}^{rm}$ are those first-order diagrams where the number of electrons entering reservoir $r$ minus those leaving reservoir $r$ is $m$.

\section{Useful relations for the $SU(N)$ generators \label{AppB}}

The generators $\lambda_a$ of $SU(N)$ fulfill the following relations
\begin{align}
\Tr (\lambda_a) &= 0 \\
\Tr(\lambda_a \lambda_b) &= 2\delta_{ab} \:, \label{EQ:prop1} \\
[\lambda_a, \lambda_b]_- &= 2i\sum_c f_{abc}\lambda_c \label{EQ:prop2} \:,\\
[\lambda_a, \lambda_b]_+ &= \frac{4}{N}\delta_{ab} + 2 \sum_c d_{abc} \lambda_c  \:. \label{EQ:prop3}
\end{align}
We can express the antisymmetric tensors $f_{abc}$ and $d_{abc}$ as
\begin{align}
d_{abc} &= \frac{1}{4}\Tr\left([\lambda_a, \lambda_b]_+\lambda_c\right) \\
f_{abc} &= -\frac{i}{4}\Tr\left([\lambda_a, \lambda_b]_-\lambda_c\right)
\end{align}
These relations will be used in the following proofs.

\section{Semi-positivity of the relaxation matrix $D^r$ \label{AppC}}
The relaxation matrix $D^r$ is defined as
\begin{align}
D^r_{ac} &= \delta_{ac} + c_N\sum_b d_{abc} n^r_b %
\:, 
\end{align}
or, equivalently,
\begin{align}
D^r \bs{g} &= \bs{g} + \bs{n}^r * \bs{g}\:.
\end{align}
We need to show that $D^r$ is positive semidefinite, i.e., $\bs{g} \cdot D^r \bs{g} \geq 0$ for any $\bs{g}$, to justify the interpretation of the corresponding term in the kinetic equation as a relaxation term.
Using $\bs{g} \cdot \bs{g} = \sum_a g_a g_a$ and $\bs{g} \cdot (\bs{n}^r * \bs{g})= c_N \sum_{abc}d_{abc}g_an_b^rg_c=c_N \sum_{abc}d_{abc}g_ag_bn_c^r$, we get

\begin{align}
\bs{g} \cdot D^r \bs{g} &=\sum_a \left\lbrace g_a g_a + c_N \sum_{bc} d_{abc} g_a g_b n_c^r
\right\rbrace \nonumber\\
&= \frac{1}{2}\sum_{ab} g_a g_b \Tr(\lambda_a \lambda_b ) \nonumber\\
&\phantom{=}+ \frac{c_N}{4} \sum_{abc} \Tr\left([\lambda_a, \lambda_b]_+\lambda_c\right) g_a g_b n_c^r \nonumber \\
&= \frac{1}{2} \Tr\left[(\bs{g}\cdot \bs{\lambda})^2\right] +  \frac{c_N}{4} \Tr\left([\bs{g}\cdot \bs{\lambda}, \bs{g}\cdot \bs{\lambda}]_+\bs{n}^r \cdot \bs{\lambda}\right) \nonumber \\
&=\frac{1}{2} \Tr\left[(\bs{g}\cdot \bs{\lambda})^2(\mathds{1}_N + c_N \bs{n}^r \cdot \bs{\lambda}) \right] \nonumber \:.
\end{align}

Since the hybridization matrix $\Upsilon^r = \Gamma^r( \mathds{1}_N+c_N \, \bs{n}^r \cdot \bs{\lambda} )$ is positive semidefinite and $\Gamma^r >0$, we can use the decomposition $\mathds{1}_N+c_N \, \bs{n}^r \cdot \bs{\lambda} = \sum_i \sigma_i \ket{i}\bra{i}$, with $\sigma_i \geq 0$. This yields
\begin{align}
\bs{g} \cdot D^r \bs{g} &= \frac{1}{2}\sum_{ij} \bra{j}(\bs{g}\cdot \bs{\lambda})^2 \sigma_i \ket{i}\braket{i|j}\nonumber\\
&= \frac{1}{2}\sum_j \sigma_j \bra{j} (\bs{g}\cdot \bs{\lambda}) (\bs{g}\cdot \bs{\lambda}) \ket{j} \nonumber \\
&= \frac{1}{2} \sum_j \sigma_j \|\bs{g}\cdot \bs{\lambda}\ket{j}\|^2 \geq 0 \:.
\end{align}
In the last line we have used the hermiticity of $\lambda_a$ and $\sigma_j \geq 0$. 

\section{$SU(N)$ invariance \label{AppD}}
Any $N\times N$ hermitian matrix $M$ can be decomposed as $M= k \mathds{1}_N + \bs{m}\cdot \bs{\lambda}$, with $k$ and $m_a \in \mathds{R}$. After a basis change $M \rightarrow \tilde{M} = U M U^{\dagger}$, we can decompose similarly $\tilde{M} = k\mathds{1}_N+  \tilde{\bs{m}}\cdot \bs{\lambda}$. The elements of $\tilde{\bs{m}}$ read: 
\begin{align}
\tilde{m}_a &= \Tr(\tilde{M}\lambda_a)/2 \nonumber \\
&= \Tr(UMU^{\dagger}\lambda_a)/2 \nonumber\\
&= \sum_{b}m_b \Tr(U\lambda_b U^{\dagger}\lambda_a)/2 \nonumber \\
&= \sum_b R(U)_{ab} m_b \:, 
\end{align}
or in matrix-vector notation $\tilde{\bs{m}} = R(U) \bs{m}$, where $R(U)$ is the $s_N$-dimensional rotation matrix corresponding to the basis transformation $U$.

The kinetic equations are written in terms of $P_0$ and $P_1$, which are obviously invariant under rotation, as well as $\bs{g}$, $\bs{n^r}$, and $\bs{B}$, which transform as vectors.
To prove the form invariance of the kinetic equation under an $SU(N)$ transformation of the basis, we need to show that the scalar product $\bs{x}\cdot \bs{y}$ transforms like a scalar and the star and wedge products $\bs{x}* \bs{y}$ and $\bs{x}\wedge \bs{y}$ like vectors.

Let us start with the invariance of the scalar product:
\begin{align}
\bs{x} \cdot \bs{y} &= \sum_{a} x_a y_a \nonumber \\
&= \frac{1}{2} \sum_{ab} x_a y_b \Tr(\lambda_a \lambda_b) \nonumber \\
&=\frac{1}{2}\Tr\left[(\bs{x} \cdot \bs{\lambda})(\bs{y} \cdot \bs{\lambda})\right]\nonumber \\
&= \frac{1}{2}\Tr\left[U(\bs{x} \cdot \bs{\lambda})U^{\dagger}U(\bs{y} \cdot \bs{\lambda})U^{\dagger}\right] \nonumber \\
&= \frac{1}{2}\Tr\left[(\tilde{\bs{x}} \cdot \bs{\lambda})(\tilde{\bs{y}} \cdot \bs{\lambda})\right]\nonumber \\
&= \tilde{\bs{x}} \cdot \tilde{\bs{y}} \:.
\end{align}

Next, we show the vector character of the star and wedge products by convincing ourselves that the combinations $(\bs{x}*\bs{y})\cdot \bs{z}$ and $(\bs{x}\wedge \bs{y})\cdot \bs{z}$ remain invariant under rotation.
We find
\begin{align}
(\bs{x} * \bs{y})\cdot \bs{z} &= c_N\sum_{abc} d_{abc} x_a y_b z_c \nonumber \\
&= \frac{c_N}{4}\sum_{abc} \Tr \left([\lambda_a, \lambda_b]_+\lambda_c\right) x_a y_b z_c \nonumber \\
&= \frac{c_N}{4}\Tr \left\{[(\bs{x}\cdot\bs{\lambda}), (\bs{y}\cdot\bs{\lambda})]_+(\bs{z}\cdot\bs{\lambda})\right\}
\nonumber\\
&= \frac{c_N}{4} \Tr \left\{[(\tilde{\bs{x}}\cdot\bs{\lambda}), (\tilde{\bs{y}}\cdot\bs{\lambda})]_+(\tilde{\bs{z}}\cdot\bs{\lambda})\right\} \nonumber \\
&= (\tilde{\bs{x}} * \tilde{\bs{y}})\cdot \tilde{\bs{z}}
\:,
\end{align}
as well as
\begin{align}
(\bs{x} \wedge \bs{y})\cdot \bs{z} &= c_N\sum_{abc} f_{abc} x_a y_b z_c \nonumber \\
&= -i\frac{c_N}{4}\sum_{abc} \Tr \left([\lambda_a, \lambda_b]_-\lambda_c\right) x_a y_b z_c \nonumber \\
&= -i\frac{c_N}{4}\Tr \left\{[(\bs{x}\cdot\bs{\lambda}), (\bs{y}\cdot\bs{\lambda})]_-(\bs{z}\cdot\bs{\lambda})\right\}
\nonumber\\
&= -i\frac{c_N}{4}\Tr \left\{[(\tilde{\bs{x}}\cdot\bs{\lambda}), (\tilde{\bs{y}}\cdot\bs{\lambda})]_-(\tilde{\bs{z}}\cdot\bs{\lambda})\right\}\nonumber \\
&= (\tilde{\bs{x}} \wedge \tilde{\bs{y}})\cdot \tilde{\bs{z}}
\:,
\end{align}
which completes the proof of the $SU(N)$ invariance of the kinetic equations.

\section{Explicit form of the Gell-Mann matrices \label{AppE}}

In the example of the triple quantum dot, we choose the standard Gell-Mann matrices for expressing the flavor-polarization vectors. These are given by:
\begin{widetext}
\begin{align}
\lambda_1 &= \begin{pmatrix}
0 & 1 & 0 \\
1 & 0 & 0 \\
0 & 0 & 0
\end{pmatrix} \qquad 
\lambda_2 = \begin{pmatrix}
0 & -i & 0 \\
i & 0 & 0 \\
0 & 0 & 0
\end{pmatrix} \qquad
\lambda_3 = \begin{pmatrix}
1 & 0 & 0 \\
0 & -1 & 0 \\
0 & 0 & 0
\end{pmatrix} \\[4pt]
\lambda_4 &= \begin{pmatrix}
0 & 0 & 1 \\
0 & 0 & 0 \\
1 & 0 & 0
\end{pmatrix} \qquad 
\lambda_5 = \begin{pmatrix}
0 & 0 & -i \\
0 & 0 & 0 \\
i & 0 & 0
\end{pmatrix} \\[4pt]
\lambda_6 &= \begin{pmatrix}
0 & 0 & 0 \\
0 & 0 & 1 \\
0 & 1 & 0
\end{pmatrix} \qquad 
\lambda_7 = \begin{pmatrix}
0 & 0 & 0 \\
0 & 0 & -i \\
0 & i & 0
\end{pmatrix} \qquad
\lambda_8 = \frac{1}{\sqrt{3}}\begin{pmatrix}
1 & 0 & 0 \\
0 & 1 & 0 \\
0 & 0 & -2
\end{pmatrix}
\end{align}
\end{widetext}

%

\end{document}